# Estimation and global control of noise reflections

Emmanuel Friot
Alexandre Gintz
Centre National de la Recherche Scientifique
Laboratoire de Mécanique et d'Acoustique
31 Chemin Joseph Aiguier
13402 Marseille Cedex 20
France

**ABSTRACT**

In theory, active control could be used to reduce the unwanted noise reflections from surfaces such as a submarine hull or the walls of an anechoic room. In the recent years, a real-time algorithm has been developed to this effect at the Laboratoire de Mécanique et d'Acoustique: the noise scattered by the surface is estimated through linear filtering of acoustic pressure signals provided by ordinary microphones and an adaptive feedforward algorithm minimizes the resulting error signals. The paper summarizes the theory underlying the control algorithm, which stems from the integral representation of the scattered pressure, and presents the successive experiments which have been conducted with it: control of terminal reflections in a duct, control of the noise scattered by a parallelepiped in an anechoic room, estimation of the noise reflections on the walls of a small room. It appears that an accurate identification of the linear filters that account for the surface scattering leads to an effective estimation and control of the scattered noise. Facilities allowing such an accurate estimation of the scattered noise are suggested for a future anechoic room where active devices would deal with the wall reflections in the 20-100Hz frequency range.

## 1. INTRODUCTION

In theory, Active Noise Control allows the suppression of scattered acoustic radiation and indeed the pioneers of ANC dedicated much effort to this subject[1], probably because during the Cold War the idea of fully invisible submarine hulls was attractive to financial supporters. However the theory which underlies these early references requires the perfect monitoring of layers of ideal sources[1], which is not easy to approximate with off-the-shelf actuators and sensors, and to the authors' knowledge this theory has not led to effective industrial applications.

A few years ago we at LMA tried to revisit the concept of scattered radiation control with the usual tools of modern ANC, i.e. multichannel feedforward control and adaptive algorithms. A real-time strategy was introduced to allow control of scattered radiation with ordinary loudspeakers and microphones. The first sections of this paper review the theory and the successive real-time experiments which have been conducted with it. It appears that an accurate off-line identification of the the so-called "diffraction filters", which maps the measured total acoustic pressure to error signals accounting for the scattered pressure linear filters, leads to an effective estimation and control of the scattered noise[2-3].

A current challenge at LMA is to build an anechoic room with embedded loudspeakers and microphones in order to cancel the noise reflections on the walls in the 20-100Hz frequency range. As measurements in a scale model of a room have shown, the identification of the diffraction filters is a poorly-conditioned inverse problem when it is conducted from measurements using an acoustic dipole, where scattered and direct acoustic fields can be easily separated; an alternative identification procedure is suggested in the paper.

## 2. A STRATEGY FOR CONTROL OF SCATTERED ACOUSTIC RADIATION

### A. Derivation of error signals accounting for the scattered radiation

The acoustic pressure scattered by any body can be classically written in the frequency domain as a function of the total acoustic pressure and pressure gradient around the body[4]:

$$p_s(\mathbf{r},\omega) = \iint_S \left[ G(\mathbf{r},\mathbf{r}_0,\omega) \frac{\partial}{\partial n_0} p(\mathbf{r}_0,\omega) - p(\mathbf{r}_0,\omega) \frac{\partial}{\partial n_0} G(\mathbf{r},\mathbf{r}_0,\omega) \right] dS_0 \quad (1)$$

where $S$ is a surface enclosing the scattering body and $G$ is the Green's Function accounting for the acoustic propagation in the medium *without* the scattering body. Earlier paper relied on equation 1 to show that adequate layers of monopole and dipole sources could cancel the scattered pressure; for usual feedforward control, equation 1 also shows that the scattered pressure can be computed by the linear filtering of the pressure and the pressure gradient around the body. Furthermore, if the pressure is given on $S$, the acoustic field inside $S$ is fully determined, except at discrete resonance frequencies when $S$ is finite, and the field is a linear function of the pressure on $S$[5]. Therefore the acoustic pressure gradient on $S$ is also a linear function of the pressure on $S$ and equation 1 can be rewritten as a function of the pressure only:

$$p_s(\mathbf{r},\omega) = \iint_S p(\mathbf{r}_0,\omega) \left[ H(\mathbf{r},\mathbf{r}_0,\omega) - \frac{\partial}{\partial n_0} G(\mathbf{r},\mathbf{r}_0,\omega) \right] dS_0 \quad (2)$$

Below some aliasing frequency, the integral in equation 2 can be approximated by a discrete sum:

$$p_s(\mathbf{r},\omega) = \sum_k g_k(\mathbf{r},\omega) p(\mathbf{r}_k,\omega) \quad (3)$$

Therefore the pressure scattered by a finite body can be estimated through linear filtering of signals from a set of ordinary microphones. Provided the filters $g_k$ in equation 3 are known, this scattered pressure can be computed at a finite set of minimization points enclosing the surface of measurement microphones and the resulting signals can be fed to any electronic active noise controller.

Numerical simulations and experiments[2,3,6] have shown that 3 measurement microphones and minimization points per wavelength allowed global control of the scattered radiation. An other experiment[7] has also shown that indeed global noise control does not work properly at the resonance frequencies of the volume enclosed by a mesh of minimization microphones. However, the theory given in this paper holds below the first resonance frequency or if more microphones are set outside the sensor surfaces enclosing the scattering body.

### B. Identification of the scattering filters

Getting the "scattering filters" $g_k$ in equation 3 is of course the key step in the process of computing error signals as suggested above. The filters are linked to the medium Green's function and to the solution of the acoustic Dirichlet problem around the body. Computing these filters from a propagation model may be very difficult in a practical case and may lead to filters that are not robust or accurate enough for estimating proper error signals; identifying the scattering filters before real-time control is a more convenient alternative. The scattering filters map a set of signals from real microphones to a set of errors signals which virtually account for scattered radiation sensors; the filters do no not depend on the noise sources, they account for the propagation medium only. Therefore they can be identified, before control of the scattered pressure due to a possibly unknown primary source, through linear inversion of acoustic data generated with the secondary sources, provided that the scattered acoustic pressure at the minimization locations can be estimated through an other process than using equation 3. Namely if $p^s_{i,j}$ denotes the scattered pressure at sensor $i$ with noise source $j$, $p_{kj}$ the total pressure at sensor $k$ and $g_{ik}$ the scattering filter accounting for the contribution of pressure sensor $k$ to scattered pressure sensor $i$, in matrix notation the scattering filters can be computed in the frequency domain from (regularized) inversion of $\mathbf{P}^s = \mathbf{G}\,\mathbf{P}$.

In order to generate the data for inversion, several distinct processes have been considered at LMA for estimating the scattered radiation $p^s_{i,j}$ at error sensors; results from experiments using these processes will be given later in the paper:
- methods involving windowing of time responses have failed, mainly because at low frequency echoes from a scattering body are very difficult to separate from actuator responses.
- If the scattering body can be removed, the scattered pressure merely amounts to the difference between acoustic pressures with and without the body. In practice this proved to be effective for getting scattered filters leading to effective real-time control of the scattered pressure in several experiments[2,3]. However, removing the scattering body from the set of sensors and actuators is not always feasible; it cannot be done in the case of an anechoic room where active control is intended to compensate for the wall reflections.
- With a perfect dipole noise sources, the acoustic pressure measured in the null plane of the dipole can only result from noise reflections; it accounts for the scattered pressure only. A dipole source has been used at LMA in a room mock-up where control is intended for canceling the wall reflections. The scattering filters could be well identified in the case of nearly rigid walls[8] but the inversion process proved to be to ill-conditioned in the case of absorbing walls.
- If secondary noise sources are located *inside* surface $S$ from equation 1, it will be shown that the scattering filters also map the total pressure on $S$ to the total pressure outside; the inverse identification of the filters is intrinsically better conditioned in this case. This suggests a specific arrangement suitable for an anechoic room which will be discussed below.

.

## C. Real-time implementation

The error signals, computed as suggested above to account for the scattered radiation, can be fed to a standard adaptive minimization algorithm for real-time control. For computational reasons it appeared[3] that the Filtered-Error Least Mean Square algorithm[9] was more suited to the minimization of the scattering error signals that the better-known Filtered-Reference LMS. If $x$ denotes the reference signal, $\mathbf{u}$ the vector of monitoring signals and $\mathbf{w}$ a set of Finite Impulse Responses, the real-time computations necessary for control can be written[3] in a form close to the usual adaptive FXLMS equations:

$$\mathbf{u}(n+1) = \mathbf{W}(n) * x(n) \qquad \text{with :} \qquad (4)$$

$$\mathbf{W}_k(n+1) = \mathbf{W}_k(n) - \alpha\,[\mathbf{H}_1 * \mathbf{p}(n-\tau+1-k) + \mathbf{H}_2 * \mathbf{u}(n-\tau+1-k)]\,x(n) \qquad (5)$$

An additional convolution product is needed when comparing these equations to the FxLMS ones; in the end the control of the scattered pressure requires about twice as much as memory and computation time that direct control of the total pressure with an FxLMS algorithm.

# 3. REAL-TIME CONTROL EXPERIMENTS

## A. 1D global control

Figure 1 depicts the first experiment that was conducted in a duct to test the control strategy introduced above. A removable tap at the duct end takes the part of the scattering body that active control has to render acoustically invisible; note that in this experiment ANC is not aimed at fully canceling the reflection at the tap but merely at recovering with the tap the acoustic field propagating without the tap, including the reflections at the open end of the duct.

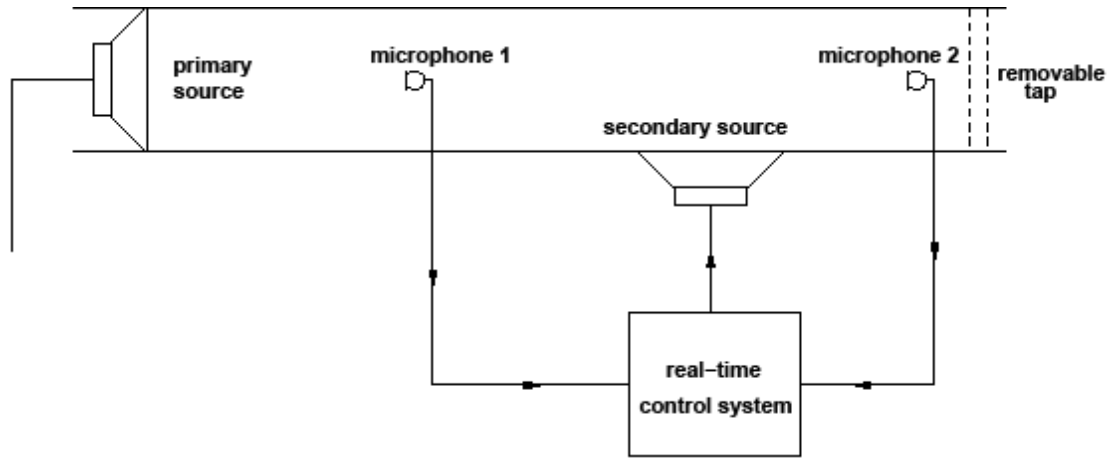

**Figure 1:** A set-up for active control of the scattered radiation due to a tap at the end of a duct.

In the set-up of figure 1, the scattered radiation by the tap can be computed by subtracting the pressure measured without the tap to the pressure with the tap. Figure 2 shows the scattered pressure with or without control when 0-100Hz white noise is monitoring the primary loudspeaker. Below 100Hz, control is not efficient because the control authority of the secondary loudspeaker is too low. Above 750Hz control is not efficient because a second acoustic mode is propagating in the duct. In the intermediate range control significantly reduces the scattered pressure. More details on this experiment can be found in reference 2.

## B. 3D partial Control

Figure 3 shows the set-up of a 3D experiment conducted in the LMA anechoic room. In this experiment active control was intended to cancel the scattered radiation by a parallelepiped at a single frequency. Since "only" 14 control channels and loudspeakers were available, global control of the scattered radiation could not be implemented. Instead numerical simulations with a Boundary Element Method suggested the arrangement of figure 3 for control of the scattered radiation at 280Hz in a restricted angular sector. Moreover, grouping the sensors and actuators in a restricted area made it easier to remove the parallelepiped in order to measure the direct acoustic pressure.

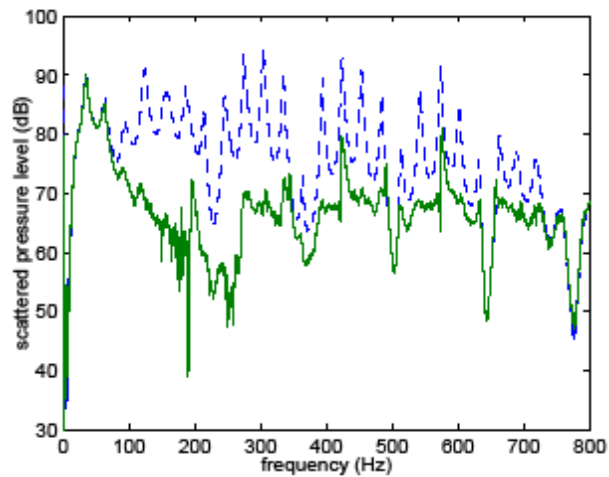

**Figure 2:** Magnitude of the scattered pressure at microphone 2 with (solid) or without (dashed) active control

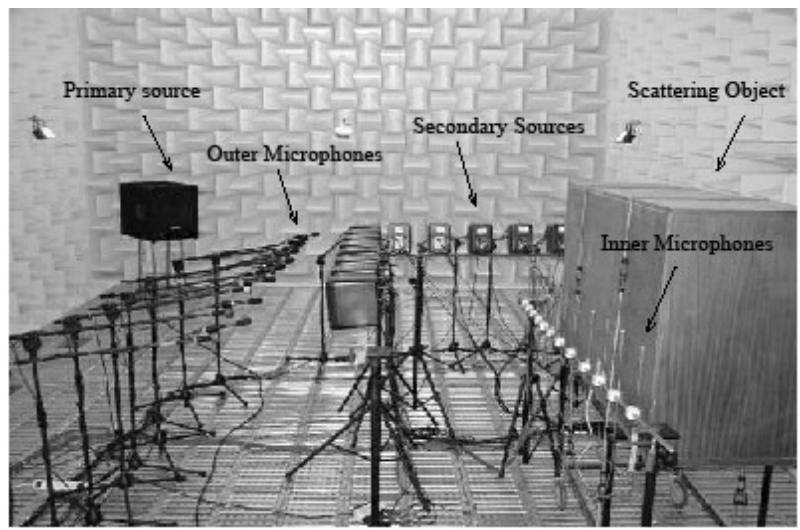

**Figure 3:** A 3D experiment of scattered diffraction control in the LMA anechoic room

Figure 4a compares the magnitude of scattered radiation by the parallelepiped as computed by subtracting to the measurements the pressure without the parallelepiped to the scattered radiation estimated with "scattering filters" as suggested in section 2. Although acoustical data for identifying the scattering filters was generate with loudspeakers in a restricted sector only, the scattered pressure is reasonably estimated with these filters. Figure 4b shows that real-time control of the error signals led in average to a 10dB reduction of the scattered pressure. More details on the numerical simulations and the experiment can be found in reference 3.

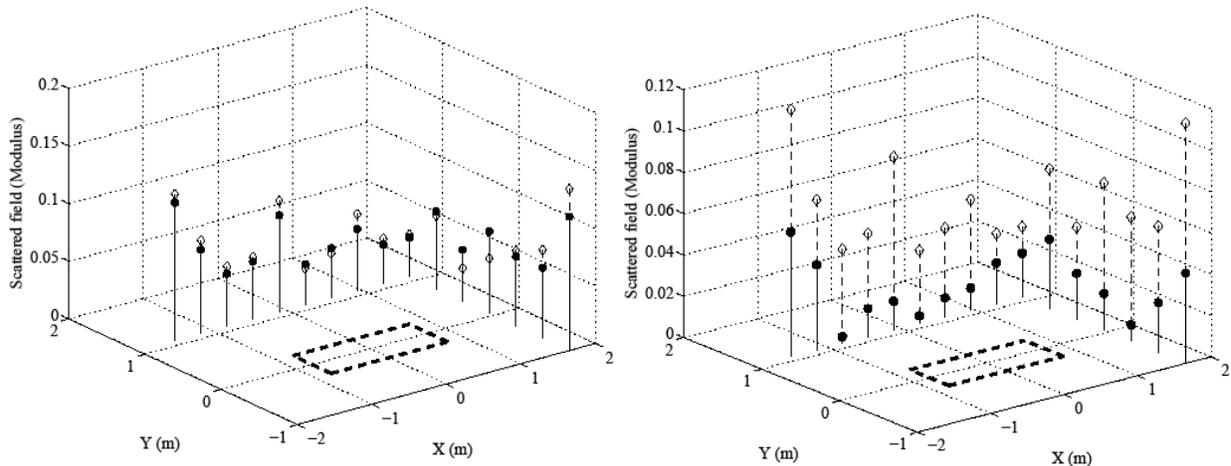

**Figure 4:** (a-left): magnitude of the measured (dashed) and computed (solid) scattered radiation by the parallelepiped with the primary noise source – (b-right): magnitude with (solid) or without real-time control

## 4. IDENTIFICATION OF SCATTERING FILTERS WITH A DIPOLE

In the latter experiment described above, active control was intended to reduce the scattered radiation around a body with a kind of arrangement that could be thought off for rendering invisible a submarine hull. However, the required number of sensors and actuators (about 3 per wavelength all around the body) makes it unrealistic to contemplate control for a real submarine with today's technology. A more sensible application of active scattering control is the reduction of the low-frequency reflections on the walls of anechoic rooms. Indeed, usual rooms do not allow anechoic measurements below 50Hz whereas such measurements would be desirable for many industrial noise sources or for validation of acoustic theoretical models.

In a 10m x 7m x 6m anechoic room such as shown in figure 3, the 3 transducers per wavelength rule suggests that about 100 microphones and loudspeakers would allow estimation and control of the wall scattered radiation up to 100Hz[6]. Although 200 transducers may seem high a number, it complies with today's controller technology and it has to be compared with the number of the room 1400 passive wedges.

It the case of an anechoic room, it is however not possible to remove the walls in order to measure the scattered radiation as in the experiments above; other processes are needed to generate data for identification of the scattering filter. At LMA A. Gintz suggested the idea of using an acoustic dipole to generate noise in a room[8,9]. The measured acoustic pressure in the dipole free-field null plane can only result from wall reflections. Therefore the room scattering filters which map the total pressure on the walls to the scattered pressure at error microphones inside the room can be identified from data generated by moving a dipole source around the error microphones.

Figure 5 shows the room mock-up that have been built at LMA to test the identification of the scattering filters with a dipole; the ceiling of the mock-up has been removed to take the picture. A mesh of 32 microphones measured the total acoustic pressure near the mock-up walls (including floor and ceiling). The dipole was rotated around an error microphone in its null plane in order to generate acoustical data for identification of the linear filters which map the total pressure at the walls to the scattered pressure at the error microphone. Figure 5 also shows the acoustic dipole which has been used in the experiments; since the mock-up is designed for the 100-300Hz range, the effective null plane of the dipole could be accurately located through measurements in the LMA anechoic room which provides free field radiation conditions above 100Hz.

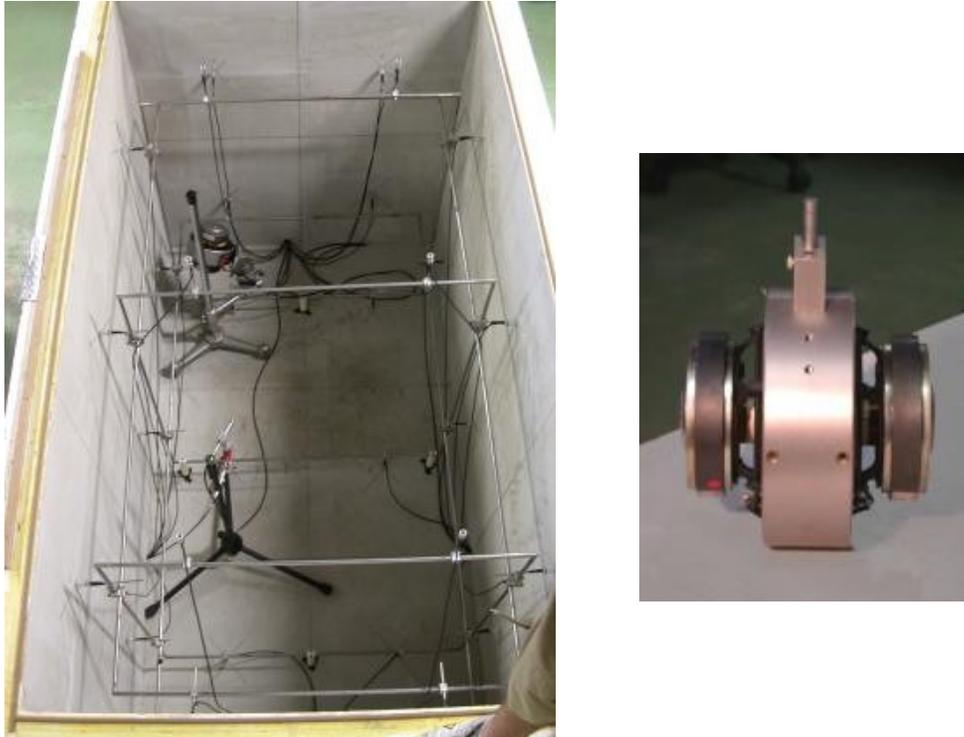

**Figure 5:** A room mock-up and a dipolar source for identification of scattering filters related to the wall reflections

Once the scattering filters have been identified for an error microphone inside the mock-up, the scattered pressure at the error microphone can be estimated whatever the orientation of the dipole is. In a case when the error microphone is on the axis dipole (and not in the null plane), figure 6 shows the measured acoustic pressure, the computed scattered and the estimated direct pressure resulting from the difference between total and scattered pressure. It appears that cavity resonances have been largely reduced in the estimated direct pressure, which indicates that the scattered field have been reasonably estimated with the identified scattering filters.

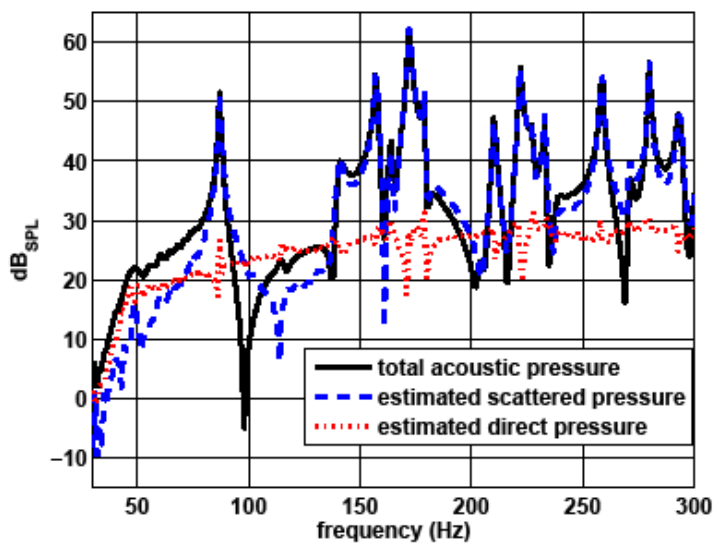

**Figure 6:** Estimation of the scattered and direct pressure from a total pressure measurement at an error microphone.

Unfortunately, similar results could not be achieved when absorbing material was laid onto the mock-up walls. With absorbing material the acoustic pressure was significantly lower at the error microphones located in the dipole null plane; the resulting inverse problem for the identification of the scattering filter was very poorly conditioned and no filters could be identified for an accurate estimation of the scattered field. In conclusion using a dipole source may not be an efficient method to identify the scattering filters in an anechoic. An other approach, which has yet to be tested in practice, is suggested in the next section.

## 5. AN ACTIVE SET-UP FOR AN ANECHOIC ROOM

The theory and experiments above suggest the arrangement of figure 7 for fitting up an anechoic room with an active system intended at reducing the low-frequency wall reflections. A primary source, whose acoustic radiation is not known, has been set-up in the center of the room; cheap detection microphones show at the surface of the absorbing material wedges; secondary loudspeakers have been mounted in the room walls.

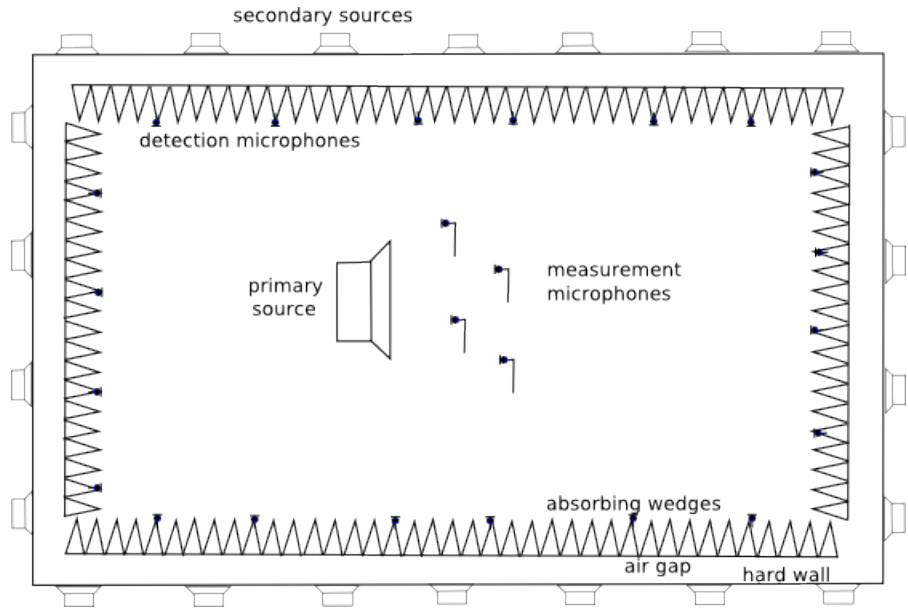

**Figure 7:** Sketch of an active setup for canceling the wall reflections in an anechoic room.

Let $S$ denote the virtual closed surface meshed by the detection microphones and $G$ the Green's function accounting for noise propagation around the primary source and microphone measurements in free space.

When the primary source is off and one or several secondary sources are emitting noise, the pressure at a measurement microphone can be written as an integral over $S$:

$$p = \iint_S \left[ G \frac{\partial}{\partial n_0} p(\mathbf{r}_0) - p(\mathbf{r}_0) \frac{\partial}{\partial n_0} G \right] dS_0 \qquad (6)$$

By comparing equations 1 and 6 it appears that the linear filters that maps, when the primary source is off, the total acoustic pressure on $S$ to the total pressure at a measurement microphone are the same as the linear filter that maps the pressure on $S$ to the scattered pressure at the measurement microphones when the primary source is on. In fact equation 6 accounts for the contribution of the whole world beyond $S$ to the pressure at the measurement microphones.

Therefore the scattering filters, that are required for estimating the scattered pressure by the room walls as introduced in section 2, can be identified by inverting measurements with the primary source off and the secondary sources on. An accurate calibration of the detection microphones is not required since the identified filters will include any gains; detection microphones are only asked to have the same response when the primary source is on as they had when it was off during excitation with the secondary sources.

Once the scattering filters have been identified from preliminary measurements with the secondary sources, feedforward control can be implemented as explained in section 2 for reduction of the scattered noise at the measurement microphones. In the set-up of figure 7 the secondary sources are located on the walls which are the source of the unwanted scattered noise, which is the better location for global control.

However, finding out an adequate reference signal for feedforward might be problematic for some primary sources such as aeroacoustic noise sources. Furthermore causality constraints may dramatically limit the performances of real-time control. In difficult cases the scattered noise may be simply computed with the scattering filters and subtracted after performing the measurements with the primary source. Even if this would be post-treatment of the data and not real-time control of the wall reflections, note that secondary sources in the walls are still necessary for the identification of the scattering filters.

## 6. CONCLUSIONS

At the time of writing, active control of the noise scattered by a submarine still sounds like an unrealistic application. However the real-time strategy which has been introduced and validated at LMA in the recent years may be suitable for active control of the low-frequency reflections on the walls of an anechoic room. A full-scale implementation of the arrangement depicted in figure 7 is now awaited.

## ACKNOWLEDGEMENTS

The authors acknowledge the Agence Nationale de la Recherche for financial support through the PARABAS (ANR-06-BLAN-0081) project. Thanks also to Claire Bordier, Nicolas Epain, Régine Guillermin, Cédric Pinhède, Alain Roure and Muriel Winninger for their support to the successive experiments, as well as to Philippe Herzog for his ever stimulating remarks.